\documentclass[manuscript,acmsmall]{acmart}

\AtBeginDocument{%
  \providecommand\BibTeX{{%
    \normalfont B\kern-0.5em{\scshape i\kern-0.25em b}\kern-0.8em\TeX}}}

\setcopyright{none}

\begin{document}
\newcommand\anedit[1]{{\color{black} #1}}

\newcommand\oldaneditb[1]{{\color{black} #1}}

\newcommand\oldanedit[1]{{\color{black} #1}}
\newcommand\angiereturnto[1]{{\color{red} #1}}

\newcommand\mklee[1]{{\color{red}{[{\bf MKL:} #1]}}}



 \title[Data and Technology for Equitable Public Administration]{Data and Technology for Equitable Public Administration: Understanding City Government Employees' Challenges and Needs}
%

\author{Angie Zhang}
\email{angie.zhang@austin.utexas.edu}
\affiliation{
  \institution{School of Information, The University of Texas at Austin}
  \country{USA}
}
\author{Madison Liao}
\authornote{The second and third authors conducted this work as a research associate at the University of Texas at Austin's School of Information.}
\email{madison.liao@utexas.edu}
\affiliation{
\institution{Department of Sociology at The University of California, Los Angeles}
\country{USA}
}


\author{Elizaveta (Lee) Kravchenko}
\authornotemark[1]
\email{kravchenko.e@northeastern.edu
}
\affiliation{
\institution{College of Arts, Media, and Design, Northeastern University}
\country{USA}
}

\author{Marshanah Taylor}
\email{marshanah.taylor@utexas.edu}
\affiliation{
\institution{School of Information, The University of Texas at Austin}
\country{USA}
}

\author{Angela Haddad}
\email{angela.haddad@utexas.edu}
\affiliation{
\institution{Department of Civil, Architectural and Environmental Engineering, The University of Texas at Austin}
\country{USA}
}

\author{Chandra Bhat}
\email{bhat@mail.utexas.edu}
\affiliation{
\institution{Department of Civil, Architectural and Environmental Engineering, The University of Texas at Austin}
\country{USA}
}

\author{S. Craig Watkins}
\email{craig.watkins@austin.utexas.edu}
\affiliation{
\institution{School of Journalism and Media, The University of Texas at Austin}
\country{USA}
}

\author{Min Kyung Lee}
\email{minkyung.lee@austin.utexas.edu}
\affiliation{
\institution{School of Information, The University of Texas at Austin}
\country{USA}
}
\renewcommand{\shortauthors}{Zhang et al.}
\begin{abstract}
City governments in the United States are increasingly pressured to adopt emerging technologies. Yet, these systems often risk biased and disparate outcomes. Scholars studying public sector technology design have converged on the need to ground these systems in the goals and organizational contexts of employees using them. We expand our understanding of employees' contexts by focusing on the \textit{equity practices} of city government employees to surface important equity considerations around public sector data and technology use. Through semi-structured interviews with thirty-six employees from ten departments of a U.S. city government, our findings reveal challenges employees face when operationalizing equity, perspectives on data needs for advancing equity goals, and the design space for acceptable government technology. We discuss what it looks like to foreground equity in data use and technology design, and considerations for how to support city government employees in operationalizing equity with and without official equity offices.

\end{abstract}




\maketitle

\section{Introduction}
U.S. cities are increasingly becoming \anedit{focal points} for purveyors of emerging government technologies---data-driven products promoted for the ability to improve efficiency and fairness of government processes. From cameras and sensors for improving traffic efficiency \cite{guerrero2018sensor} to algorithms for automating bureaucratic processes or allocating limited resources more effectively \cite{janssen2016challenges, sorensen2018algorithms, henman2020improving, toros2018prioritizing, chouldechova2018case}---governments are increasingly exploring, or being solicited to explore, tech products. For institutions whose aim is to deliver economic, effective, efficient, and equitable services to constituents, the promises of such products are \anedit{appealing}. 

However, past work has documented that government use of technology and data-driven processes walks a fine line between promoting accessibility and efficiency, and exacerbating inequality and marginalization \cite{eubanks2018automating, borry2019automation, william2012public, o2017weapons, rainie2017code, lorinc2022dream}. Instances of harms aggravated by technology include increasing marginalization due to the growing digital divide \cite{ebbers2016impact, helbig2009understanding, van2020digital}; intensifying surveillance, censorship, and targeting practices \cite{usaid_2022, brayne2014surveillance, hayes2012surveillance, mozur2019one}; reproducing human bias and inequality through automated systems \cite{eubanks2018automating, usaid_2022, kuziemski2020ai, lyon2003surveillance, cheng2022child, brayne2014surveillance}; and expanding data privacy concerns \cite{ismagilova2020security, green2017open, cui2018security}.

One approach \anedit{human-computer interaction} (HCI) and \anedit{computer-supported cooperative work} (CSCW) scholars have used to probe how harmful outcomes are engendered and how to mitigate them is investigating the perspectives of public administration employees using or developing such systems and technologies, particularly algorithms and automated decision-making systems \cite{holten2020shifting, veale2018fairness, lenhart2022collect, kawakami2024situate, kawakami2024studying}. This work has surfaced that problems can arise from 1) a misalignment between the embedded values \oldaneditb{and} goals of technology with those of employees \cite{veale2018fairness, holten2020shifting, saxena2021framework}, 2) missing context that only end-users (e.g., employees) or impacted users (e.g., constituents) can provide \cite{kawakami2022improving, saxena2021framework, kawakami2022care, cheng2022child, kim2024public}, and 3) bureaucratic impediments employees contend with such as budgetary constraints favoring economy over equity when designing or selecting tools \cite{saxena2021framework, veale2018fairness, kawakami2024situate, kawakami2024studying}. 

\oldaneditb{This body of work surfaces a growing need to understand public administration employees' perspectives situated in their contexts---e.g., values they operate by, roles and organizational dynamics.} \textbf{Thus, we propose expanding research on public administration technology and data use by turning our attention to employees' \textit{equity} practices and goals \anedit{that have} been a lesser focus of past work.} We are motivated by the longstanding pursuit of equity by public administration \cite{frederickson1990public, blessett2019social} and the rise of equity initiatives in cities \cite{national_league_of_cities_2022}. Since the 1960s, equity---\oldaneditb{the pursuit of "fair, just and equitable management" when serving the public \cite{ASPA}}---has been upheld as a pillar for public administration employees to seek. Yet, employees continue to face practical challenges and trade-offs when striving towards equitable outcomes for citizens, including the adverse impacts technology and data-driven processes have had on equity goals \cite{ruijer2023social}. \oldaneditb{Additionally, in the U.S., a number of city governments are initiating or expanding equity initiatives. These are efforts that} include the establishment of equity offices, creation of equity training for government employees, and passage of ordinances to support equitable allocation and delivery of public services \cite{national_league_of_cities_2022}. \oldaneditb{Presently though, these equity initiatives have been less explored, especially around how employees are trying to pursue them and what the resulting impacts are.} \oldaneditb{Explicitly focusing on how employees pursue equity in the day to day} can help us 1) understand the equity work practices and challenges employees face, and 2) surface design implications that respect their organizational conditions and constraints. To that aim, we ask the following research questions: 

\begin{quote}
    \item \textbf{RQ1:} What are city government employees' equity goals?
    \item \textbf{RQ2:} What are their work practices and challenges when pursuing equity goals?
    \item \textbf{RQ3:} What are the data and technology opportunities and challenges that they identify for advancing equity goals?
\end{quote}

To answer these questions, we conducted semi-structured interviews with thirty-six employees from ten departments of a local city government.  Probing their equity goals \textbf{(RQ1)}, we find that at first glance, these seem straightforward, from internal equity goals related to government hiring practices and performance reviews to external equity goals of providing targeted assistance to citizens in need. However, a closer look into their work practices \textbf{(RQ2)} reveals challenges employees face when operationalizing equity, including misalignments in defining equity and weakened accountability due to a lack of formalized roles and responsibilities. Then, we present the data needs employees raised as shaped by equity challenges and considerations, followed by the design space they identified for acceptable technology \textbf{(RQ3)}. Based on our findings, we discuss what it looks like to foreground equity in data use and technology design and how to support city government employees towards meaningful use of tech and data in the context of equity. 

\section{Background on Public Sector Technology}
We begin by describing the growing interest in public sector technology and data-driven processes, and the real-world harms that have resulted. We then summarize how HCI and CSCW scholars have sought to understand and address these harms, describing the design gaps they have identified as well as the methodologies researchers are exploring to address these gaps.

\subsection{Growing Interest in Expanding Public Sector Technology and the Resulting Inequities}\label{background_grow}

Historically, governments have been slow to adopt new technology, such as early tools for delivering information and services to the public through information and communications technology (ICT) \cite{weerakkody2005exploring, conklin2007barriers}. Over time though, they have increased their technology and data-driven processes. \oldaneditb{This includes} digitization of hand-written forms, to automation of interactions between government employees and the public (e.g., online payments) \cite{bovens2002street}, and more sophisticated tools such as the use of artificial intelligence (AI) to generate predictions or assessments \cite{young2019artificial, veale2019administration, hinkley2023technology}. At the time, and even now, proponents of increased technology adoption \anedit{continue to tout} the potential for improved efficiency \oldaneditb{(e.g.,} lower processing times\oldaneditb{)} and economy \oldaneditb{(e.g.,} lower costs\oldaneditb{)} for the public \cite{national2002information, von2004electronic, archmann2010egovernment, merhi2020effective}. 

However, this growth in government technology use and data-driven processes for economy and efficiency has also been accompanied by evidence of mixed impacts on \textit{equitable} distribution and access to services, as well as new concerns such as privacy fears over heavy public surveillance for data collection \cite{brayne2014surveillance, hayes2012surveillance, mozur2019one, ruijer2023social}. For example, data-driven technologies that rely on large amounts of data to produce targeted assistance (e.g., for social work) have been criticized for over-surveilling and stigmatizing low-income or minority populations \cite{dobson2019welfare, madden2017privacy}, and amplifying past inequities rooted in human decision-making \cite{eubanks2018automating, shapiro2017reform}. Even technologies such as civic tech intended to improve public participation and procedural justice in government processes show evidence of increasing inequitable public service delivery due to varying levels of trust, access to government, and knowledge about how to participate across different communities \cite{xu2020closing, dickinson2019cavalry}. Furthermore, not only do governments face pressures to be more data-driven for efficiency and transparency purposes \cite{van2019data, matheus2020data, kim2024public}, they must also contend with backlash when new technology implementations go awry \cite{ho2023child, ghaffary2019new}, upskilling challenges from the rapid advancement of tech product capabilities \cite{Deloitte_2023}, and continued overpromises by vendors about their products’ benefits \cite{lorinc2022dream, johnson2015web}. 

\oldaneditb{This body of work suggests that when implementing data and technology, competing public administration priorities---especially economy and efficiency---often take precedence and risk inequitable outcomes. 
Our study furthers an understanding about these competing priorities by contextualizing how employees of a local city government try to operationalize equity goals.}



\subsection{Investigating Public Sector Technology and its Shortcomings} \label{background_investigate}
One line of inquiry that CSCW and HCI researchers have pursued towards equitable outcomes from government technology and data use has been investigating public sector algorithms, especially algorithmic decision-making systems (ADS), and the perspectives of the employees using or developing such systems \cite{holten2020shifting, veale2018fairness, lenhart2022collect, paakkonen2020bureaucracy, alkhatib2019street, kawakami2022improving, saxena2021framework, kawakami2022care, cheng2022child, de2020case, saxena2022train, gondimalla2024aligning}. \oldaneditb{Uniquely, Kim et al. \cite{kim2024public, kim2024integrating} also engaged with local community stakeholders who interact with the government's building services (e.g., small business owners navigating complex permit applications to open their storefronts).} 
These studies highlight how inequitable outcomes \oldaneditb{can be} produced: end-users (e.g., employees) and impacted constituents (e.g., citizens) \oldaneditb{can reveal critical} context\oldaneditb{ual factors to inform responsible technology design \cite{kawakami2022improving, saxena2021framework, kawakami2022care, cheng2022child, saxena2022train, kim2024public, kim2024integrating, kim2024public, kim2024integrating}. By not engaging with these groups, AI developers risk creating systems that are imbued with misaligned values and goals.} 
%
%
%
Factors such as bureaucratic resource constraints \cite{saxena2024algorithmic, robertson2021modeling} and unaccounted-for organizational norms and dynamics \cite{kawakami2024studying, lenhart2022collect, kawakami2022care} can also contribute conflicts between the expectations on which technology systems are built and how they work in reality. 

To bridge this gap, some researchers have pursued design research activities with employees \oldaneditb{or} impacted users of systems \cite{kawakami2022improving, kuo2023understanding, gondimalla2024aligning, scott2022algorithmic} to surface the functionalities desired from public sector tech towards equitable outcomes. Researchers have also proposed frameworks and resources (e.g., \cite{saxena2021framework, kawakami2024situate, nelson_toolkit_2020, krafft2021action}) to support designing or evaluating public sector technology. Still others have offered lenses of inquiry from domains outside of HCI to expand our understanding of public sector contexts, e.g., public administration theories \cite{alkhatib2019street, saxena2021framework, young2019artificial}, legal perspectives \cite{lyons2021conceptualising}, science and technology studies (STS) concepts \cite{saxena2021framework}, and anthropological methods \cite{kawakami2024studying}. For example, \citet{alkhatib2019street}'s use of public administration's street-level bureaucracy theory to analogize sociotechnical algorithms has motivated CSCW researchers to investigate how to design algorithmic systems to responsibly support "human discretion"---the ability of bureaucrats to make decisions "(in which uncertainty is present) for a complex task about how best to deliver government services, benefits, and punishments" \cite{bullock2019artificial}. \citet{saxena2021framework} similarly draw from public administration and STS scholarship to propose a framework for designing and evaluating public sector algorithms. Recently, \citet{kawakami2024studying} proposed using the anthropological lens of "studying up" to characterize the power dynamics that impact how government employees of different roles can influence AI decision-making. Drawing from these different fields helps illuminate employees' perspectives and contexts that public sector algorithms and technology should align with.

We believe a valuable complement to \oldaneditb{how CSCW scholars have studied} public sector data and technology \oldaneditb{use and design is by focusing on equity contexts} grounded in public administration scholarship to probe the work practices of government employees. \oldaneditb{By doing so, our study aims to make visible employees' equity goals, their work practices and challenges when pursuing these goals, and related data and technology opportunities and challenges.}

\section{Our Approach: Studying the Public Administration Effort for Equity}

\oldaneditb{To inform the goals of our research questions---what are city government employees' equity goals, related work practices and challenges, and potential opportunities and challenges related to data and technology---}we situate how public administration has grappled with operationalizing equity. 

"Social equity" first arose as a key principle for public administration to pursue in 1968 \cite{gooden2011advancing}. Similar to how a number of equity initiatives were introduced following the racial unrest in 2020 \cite{Americanbarorg_2021}, public administration scholars in the sixties were wrestling with how to respond to civil and social unrest---particularly around \textbf{racial inequality and injustice}---and held the now seminal 1968 Minnowbrook Conference to address the need for institutional change \cite{frederickson2005state, wooldridge2009epic}. Here, \citet{frederickson1980new} asserted the need to actively pursue \textbf{social equity} in addition to traditional objectives of public administration---efficiency, offering "more or better services with available resources", and economy, maintaining the "level of services by spending less money". While he did not offer a strict definition of social equity, he urged administrations to pursue it through "activities designed to enhance the political power and economic well being of these [disadvantaged] minorities" \cite{frederickson1980new}. Social equity was further institutionalized as a value of public administration in 2005 when the National Academy of Public Administration (NAPA) declared it a pillar of public administration alongside efficiency, effectiveness, and economy \cite{NAPA_2020}. 

Despite this formalization, carrying out social equity has been \anedit{far from} straightforward. Just as we documented in \ref{background_grow} how public sector technology has been designed for lower costs and higher throughput at the expense of equity, public administration employees have often also prioritized efficiency over equity as the former can lead to more immediate payoff than the latter \cite{norman2011balancing, cepiku2021equity, redden2018democratic, fernandez2019equity}. Also, public administration scholars are generally aligned that the theoretical definition of equity encompasses fairness---providing "due process, equal protection, and equal rights to all persons regardless of their personal characteristics" \cite{svara2005social}---and justice---that "all are treated fairly and get what they deserve" and "individuals can obtain a remedy if freedom or equality is denied" \cite{johnson2015justice}. However, as social equity has expanded to multiple dimensions beyond earlier considerations of race and ethnicity---e.g., gender, income, education \cite{frederickson2005state, svara2005social, gooden2011advancing}---attempting to satisfy equitable outcomes for all dimensions has often proven difficult.

Additionally, though some scholars claimed that racial inequities have abated \cite{shafritz1997public, frederickson2005state}, a closer look at outcomes adjusted for other social equity characteristics---e.g., education and income---shows that inequities along racial lines persist, such as within healthcare \cite{nelson2002unequal} and food insecurity \cite{bowen2021structural}. In fact, researchers have shown that racism continues to be exacerbated by U.S. public administration, such as through the racialized use of police force \cite{pandey2022new} and divestment of water treatment facilities in mostly Black, metro areas \cite{house2021institutional}. Thus, some scholars have pushed to place \textbf{racial equity} (back) at the forefront of U.S. public administration efforts as a first step for any meaningful social equity progress \cite{gooden2015race, lopez2018advancing, moloney2023flawed, blessett2018embedding, house2021institutional, pandey2022reckoning, pandey2022new, riccucci2021applying}. But even this narrower definition of equity has been a challenge to undertake as scholars like \citet{gooden2015race} have shown that in practice, bureaucrats often exhibit a "nervousness" to directly address race and equity.

As shown above, equity as a pillar of public administration has been a tangled pursuit for U.S. governments. \oldaneditb{Accordingly}, we position that \oldaneditb{centering the complex nature of practicing equity can offer} meaningful \oldaneditb{insights for CSCW researchers interested in shaping equitable public sector technologies}. By investigating how employees practice and pursue equity, we can illuminate important perspectives and challenges they contend with to inform implications for meaningful public sector technology and data use. \oldaneditb{To our understanding, such a qualitative study focusing on the equity practices of employees in context to \anedit{surface} public sector data and technology implications has not yet been undertaken.} 
\newline



\section{Method}
\subsection{Background: Equity Practices in City}
We spoke with employees of one of the top 15 largest cities in the U.S.\footnote{We refer to the city as "City" throughout the paper.}, which has grappled with and acknowledges a history of racism and legacies of systemic inequities that extend into present day. 
In 2016, in response to a community coalition that demonstrated how \anedit{numerous} health, economic, and quality of life disparities in City were largely determined by race, the Equity Office was created to promote equity in City's internal and external operations. Due to this unique origin, the Equity Office took the lead from the community to define equity such that "\textbf{race} is no longer an indicator of one’s quality of life".
To execute their mission, the Equity Office was tasked with conducting equity assessments with each department to evaluate existing practices, policies, and impact on racial equity, as well as provide recommendations for improvement.\footnote{\oldaneditb{The department-level equity assessments conducted by the Equity Office are separate of the study described in this paper.}} 
%


%
    
\subsection{Participants}
We held two rounds of interviews between Summer 2021 and Winter 2022. 
The first round was to get a broad understanding of employee data and technology practices, \oldaneditb{focused on} equity. The second round was held to hear participant feedback on our interpretations of the first round of interviews and \oldaneditb{their equity goal progress---}successes or challenges. In total, we spoke to thirty-six city employees from ten departments. 

The Equity Office helped us identify and schedule first round interviews by distributing an email to department heads to solicit participation in interviews exploring technology use, \oldaneditb{with a specific focus on} equity. \oldaneditb{Interested in learning about different departments' data and technology practices, the Equity Office asked to attend interviews as well.}
In the first round, we spoke to twenty-two employees from nine different departments. Interviewees included employees in tech-related roles---such as IT Supervisor or Manager, IT System Architect, and Data Manager, as well as less technical roles who utilize or implement technological solutions for their department---such as Marketing Manager, Program Manager, and Business Process Consultant. 

\oldaneditb{Analyzing these interviews, we realized potential biases \anedit{could arise} due to the Equity Office's involvement. \anedit{In City, the Equity Office and departments hold equal hierarchical standing in the organization. However, we wondered if the Equity Office's charge to conduct equity assessments with each department could influence a power dynamic favoring the office.} For example, because the Equity Office sent the emails soliciting participation in first round interviews, perhaps participants felt pressure to participate or answer a certain way due to concerns it could reflect poorly otherwise. Thus, when arranging second round interviews, we reached out to past participants directly to minimize potential bias from involvement of the Equity Office.} 
%
For these, we separately contacted past participants and asked if there were others in equity positions---employees in roles or performing tasks related to equity in their respective departments---that they wished to include or refer us to. We also reached out to the Equity Office for their reflections on the first round interviews. We spoke to nineteen employees from the nine departments that responded. \anedit{From these, we learned nuances that shed new light on the relationship and power dynamic between the Equity Office and departments, elaborated in Section \ref{findings_subsection_eo}.}

The departments we spoke to include: Transportation (P1-3), Development (P4-6), Health (P7-10), Housing (P11-15), Fire (P16), City Internal Technology (CIT) (P17-21), Watershed (P22-26), Small Business (P27-31), Innovation (P32-35), and Equity (P36). This is summarized in Table \ref{table:interviews}. 
\begin{table}[htbp]
\begin{tabular*}{\linewidth}{@{}cccc@{}}
\toprule

\multicolumn{1}{c}{\begin{tabular}[c]{@{}c@{}}\textbf{Department \&} \\ \textbf{Related Services} \end{tabular} 
}
&\multicolumn{1}{c}{\begin{tabular}[c]{@{}c@{}}\textbf{Total \# of} \\ \textbf{Participants} \end{tabular} 
}

&\multicolumn{1}{c}{\textbf{First Round}} 
&\multicolumn{1}{c}{\textbf{Second Round}}

\\ 

\midrule

\oldaneditb{\begin{tabular}[c]{@{}c@{}}\textbf{Transportation} \\ Manage mobility \& safety of\\City's transportation systems \end{tabular} }
& 3 & P1, P2, P3 & -  \\ \midrule

\oldaneditb{\begin{tabular}[c]{@{}c@{}}\textbf{Development} \\ Oversee development processes: \\e.g., permits, inspections, code compliance \end{tabular}} & 3 & P6 & P4, P5, P6  \\ \midrule

\oldaneditb{\begin{tabular}[c]{@{}c@{}}\textbf{Health} \\ Educate \& assist around \\health, shelter, food, clothing, jobs \end{tabular}}  & 4 & P7, P8, P10 & P8, P9  \\ \midrule

\oldaneditb{\begin{tabular}[c]{@{}c@{}}\textbf{Housing} \\ Support \& expand affordable \\housing opportunities \end{tabular}} & 5 & P13, P14, P15 & P11, P12  \\ \midrule

\oldaneditb{\begin{tabular}[c]{@{}c@{}}\textbf{Fire} \\ Protect \& educate community on  fires,\\e.g., firefighting, code enforcement \end{tabular}} & 1 & P16 & -  \\ \midrule

\oldaneditb{\begin{tabular}[c]{@{}c@{}}\textbf{CIT} \\ Support City's department-level\\ IT operations and needs \end{tabular}} & 5 & P19, P20, P21 & P17, P18  \\ \midrule

\oldaneditb{\begin{tabular}[c]{@{}c@{}}\textbf{Watershed*} \\ Protect citizens from flood, \\erosion, \& water pollution risks \end{tabular}} & 5 & P25, P26 & \begin{tabular}[c]{@{}c@{}}P22, P23, P24,  \\ P25, P26 \end{tabular}  \\ \midrule

\oldaneditb{\begin{tabular}[c]{@{}c@{}}\textbf{Small Business} \\ Provide training, educational events, \\\& coaching for small businesses \end{tabular}} & 5 & \begin{tabular}[c]{@{}c@{}}P27, P28, P29,  \\ P30, P31 \end{tabular} & P27  \\ \midrule
 
\oldaneditb{\begin{tabular}[c]{@{}c@{}}\textbf{Innovation} \\ Pursue research \& design initiatives with \\City departments \& community partners \end{tabular}} & 4 & P35 & P32, P33, P34   \\ \midrule

\oldaneditb{\begin{tabular}[c]{@{}c@{}}\textbf{Equity} \\ Set City's equity mission \& \\guide departments on equity initiatives \end{tabular}} & 1 & - & P36  \\ \midrule

\end{tabular*}
\vspace{2mm}
\captionsetup{width=1\textwidth}
\caption{Participant table of departments and employees we spoke to for each round of interviews. \\ *\oldaneditb{City has separate departments that manage public utilities such as water, wastewater, and electricity.}}
\label{table:interviews}
\end{table}

\subsection{Interview Protocol, Procedure, and Analysis}
The first round of interviews focused on understanding employees' overall data and technology practices and whether \oldaneditb{and} how they incorporate equity considerations. We asked employees to describe their department’s equity goals, the technology systems they use, how these are procured, and technology policies and operations. We explored how departments evaluate the outcomes of technology use (\textit{"How do these tools impact your ability to serve city residents equitably?"}) and whether procedures exist to evaluate systems for biases or issues. Final questions probed how departments engage the public around technology-related matters (\textit{"What procedures does your department have in place to engage community members about technology related matters?"}).

The second round of interviews focused on validating the accuracy of our interpretation of the first round of interviews \cite{creswell2000determining} and asking additional questions for further insights. We began by sharing takeaways from each department's first round interview to give participants the opportunity to confirm, dispute, or clarify points. We then asked employees to explain how their department defines and operationalizes equity, how employees individually operationalize equity, and inquired about equity updates in the department (\textit{"Are there other changes your department has implemented as it relates to supporting equity?"}). We ended with their thoughts for future data and technology they support or oppose for their department (\textit{"Is there any AI that you would like your department (not) to be using?"}).

Interviewers of first round interviews included our research team and a representative of the Equity Office. Interviewers of second round interviews consisted of research team members. Interviews were held per department. Both rounds of interviews lasted 1-1.5 hours with 1-5 employees from their respective departments, the number of attendees varying due to availability. They were conducted and recorded using Microsoft Teams or Zoom and transcribed using Otter.ai. All data was stored on University cloud software, accessible only by the research team.

Separate analysis of first and second round interviews followed a qualitative data analysis method \cite{patton2014qualitative}. Three researchers coded the transcripts at the sentence or paragraph level using Dedoose, a qualitative analysis software, in a deductive-inductive manner: parent codes were predefined based on interview sections, e.g., \textit{"Equity goals"}, and child codes were created inductively off these, e.g., \textit{"Past shortcomings in addressing equity"}. \oldanedit{Emerging themes and takeaways were discussed in bi-weekly team meetings.} Two researchers then transferred codes \oldanedit{and corresponding quotes of both interview rounds} to Miro, a collaborative whiteboard application, to visualize emerging themes. \oldanedit{Codes of both interview rounds were combined using axial coding \cite{williams2019art} to obtain four high-level themes: "Operationalizing Equity", "Equity Context", "Data and Equity", and "Tech and Equity". 
}

Our research team obtained City employees' consent at the start of interviews, emphasized participation was voluntary, and explained that interview data would be anonymized and kept in a secure database only accessible by our academic research team. 


\subsection{Researcher Positionality \& Research Ethics} 
Our research team is made up of undergraduate students, graduate students, and faculty. Members come from South Asian, East Asian, African American, Middle Eastern, and Caucasian ethnicities. Team members identify as female, male, and non-binary; from academic backgrounds in Transportation, Information, Sociology, English, Journalism, and Human-Computer Interaction; with experiences in quantitative and qualitative research methods. No members hold official appointments at City or have conflicts of interest with City. All members resided in City at the time of interviews and therefore have a degree of personal investment in City's equity practices. %

\oldaneditb{We reflect that} our team's diversity influenced \oldaneditb{our research and analysis in three significant ways. First,} the interdisciplinary nature of our team shaped our decision to start by understanding employees' equity context from a non-technological standpoint \textit{before} introducing data and technology factors. \oldaneditb{Second, team members with interests in critical methods helped us recognize the importance of conducting follow-ups without Equity Office employees to give participants space to speak more openly and share feedback on our first round interview interpretations \cite{creswell2000determining}.} \oldaneditb{Third, our team wrestled with how to examine employees' interpretation of equity---racial equity in particular---and analyze this in our findings. In group meetings to discuss emerging interview themes, we often also shared our personal experiences with racial discrimination and equity, and interactions with our government representatives, such as navigating complex permitting applications. In fact, all but one of our team members come from a community of color, and many of us are first- or second-generation immigrants. 
}

Though we worked together with City employees to conduct first round interviews, our research team analyzed all interview results independently. Findings and drafts of our manuscript were shared with the employees we interviewed. 

\section{Findings} 
\anedit{To address RQ1 \& RQ2, w}e first describe participants’ equity context \anedit{and related challenges}: how they define equity, their equity goals, and how they operationalize equity (\ref{Findings_Context}). \anedit{For example,} employees struggle to operationalize equity due to factors such as (1) \anedit{strong predispositions to pursuing "equality"} and (2) the degree to which equity roles and responsibilities are formalized. 
\anedit{To address RQ3, w}e then describe employees' data practices and needs shaped by equity considerations (\ref{Findings_data}), followed by their perceptions about what are (not) appropriate uses of technology by City government (\ref{Findings_tech}).
Notably, when participants described data needs, they \textit{did} center \anedit{pursuing} equity over equality and exhibited awareness of data trade-offs, \anedit{such as granular data collection enabling \anedit{measurement of} equity goals but also risking misuse}. \anedit{Similarly}, when sharing ideas for public sector technology, participants described these in connection to equity impacts not only pertaining to City's community members but also City's government employees.

\subsection{Establishing the Equity Context of City's Government}\label{Findings_Context}
\oldaneditb{Probing employees' equity goals and pursuits helped us gain a sense of how tensions arise when they are trying to translate equity \anedit{from theory} into practice. First, though employees aligned on how they define equity, some seemed resistant or nervous to shift away from equality \anedit{to} achieve equity. Next, different departments held similar equity goals, such as ensuring internal equity in hiring practices and providing targeted assistance to underrepresented communities. Yet, they varied in how much clarity they felt they had on equity directives and their personal accountability, often related to differences across departments of formalized equity roles and responsibilities. Finally, we were surprised to learn that despite their formal title, the Equity Office actually has limited ability to enforce departments and employees upholding and pursuing equity goals. This was helpful background about organizational and equity complexities at play prior to introducing factors of data and technology.}

\subsubsection{\textbf{Defining equity: Employees exhibit lingering challenges differentiating equity from equality and a tentativeness around race.}}
The Equity Office defines equity to be where "\textbf{race} is no longer an indicator of one’s quality of life". Most employees confirmed alignment with City’s definition\footnote{in both rounds of interviews, including second round interviews without the presence of Equity Office employees}. However, a couple employees spoke instead about different facets of equity, such as P18 who views race as a subset of equity that should also encompass other attributes including gender, age, and pregnancy status. Additionally, \textbf{discerning between equity and equality was a challenge.} Some expressed that equity is "not necessarily fairness but outcomes" (P5) and practicing equity may mean expending more resources to help one person reach the same desired outcome as another (P9); others described equity as "equal fairness" (P6) or "treat[ing] everybody the same" (P16). 

P36 from the Equity Office observed that some employees have difficulty shifting away from equality, attributing this fixation to a belief that public servants should "operate from a colorblind perspective". This has been documented in public administration scholarship as an ongoing challenge for practitioners to distinguish between equality and equity when pursuing the latter \cite{gooden2015equality, gooden2015race}. It is also reminiscent of AI fairness approaches for mitigating biases via "fairness through unawareness"---the exclusion of sensitive attributes when training classifiers \cite{wang2021fair, grgic2016case}. 

\subsubsection{\textbf{Equity goals: Department high-level objectives are aligned but organizational support and norms for operationalizing equity differ across departments.}} Employees of all departments described similar overarching equity goals. First, they have internal equity goals centered on raising City employee awareness around racial equity and ensuring equity in hiring practices, performance reviews, and internal resource allocation. They also hold external equity goals which focus on understanding gaps for whom departmental services are reaching. In the short-term, employees aim to provide targeted assistance to those in need, and in the long-term, right the wrongs of systemic policies. This distinction of short-term versus long-term was notable and re-emerged when several participants expressed desires for data to support measuring and analyzing long-term equity (\ref{Findings_data}).

Though their overall equity goals were similar, we observed across departments a difference in equity awareness, accountability, and operationalization, often connected to the degree to which equity roles and responsibilities had been formalized, explained next.

\subsubsection{\textbf{Operationalizing equity within departments: Formal roles and responsibilities strengthen cooperation, awareness, and clarity on equity directives, while the absence leads to confusion on individual accountability and \oldaneditb{reliance on voluntary efforts to advance equity}.}} In departments where equity roles and responsibilities are formalized, participants reflected having clear directives about how to pursue equity. For example, of all the departments, Health shared having the most official equity positions with the clearest objectives: two internal equity roles---an HR advisor \oldanedit{to review hiring practices} and a training specialist \oldanedit{to develop City's first formal department-wide equity training}, and three external equity roles---an equity and inclusion program manager to oversee all equity related efforts, a community engagement specialist to evaluate community outcomes for gaps, and a business process specialist to assess community health data and identify action items. Even for non-equity specific roles, \oldaneditb{P9 explained that} Health is building out formal ways to legitimize equity work: employees who take on such tasks can earn up to 8 hours of PTO to allow them to "tap out" and recharge.

\textbf{On the other hand, departments with no official equity positions, or equity positions that lack explicit charges, described relying on informal \anedit{approaches} or optional activities to advance equity.} Employees in equity roles explained working to normalize equity as a lens for their departments, such as offering equity trainings employees can (voluntarily) take. They also described engaging in activities to learn more about their departments' charges: P5 explained that soon, he and P4 would shadow City inspectors working face-to-face with the community "so we can see what our inspection staff encounters when they go out into the world" and inform future equity goals and training content. Uniquely, Watershed's first equity program manager (P23) explained she was spending her first six months learning the department's history with City's citizens in order to create an equity charter for Watershed. This intentionality in Watershed for the equity manager to study the department's history with citizens was also reflected in a high equity awareness when employees later talked about desires for granular data to evaluate for racial equities (\ref{Findings_data}). 

Employees not in explicit equity roles exhibited varying levels of equity awareness and involvement. While some like P8 and P18 were self-motivated to serve as equity ambassadors or on department equity boards, and others in departments such as Small Business and Innovation expressed equity as inherently embedded in their responsibilities ("That [operationalizing equity] is our day to day work." ---P32 from Innovation), \textbf{P22 of Watershed reflected how their data scientists "just want to be told what to do" because they are nervous about messing up.} 

\subsubsection{\textbf{Operationalizing equity within the Equity Office: The office holds limited authority and turns to alternative strategies to pursue instead.}} \label{findings_subsection_eo}
Despite being City's equity leads, \oldaneditb{P36 told us} the Equity Office faces limitations in exerting wide-scale influence, \oldaneditb{attributing} much of this to three factors. First, \oldaneditb{he explained that} being an office of ten serving \oldaneditb{City's workforce of} over ten thousand \oldaneditb{employees} constrains their ability to scale up training and equity assessments. Second, \textbf{the office lacks official power to compel departments to take action.} \oldaneditb{P36 was frustrated over how} administering the equity assessment is \oldaneditb{the office's} most far-reaching authority, yet departments can delay these without consequence and following resulting recommendations is optional. Still, P36 was conscientious that making actions mandatory risks 1) jading employees on the importance of equity and 2) misleading employees that undergoing one equity assessment is a seal of approval in perpetuity, when in reality, equity must be evaluated continuously. \textbf{Third, there is a culture of apprehension within City about working with a racial equity focus:} "Even now, we will still hear...fear, and this feeling that we're not even supposed to talk about race, or that we can't talk about race" (P36), \oldanedit{recalling \citet{gooden2015equality} describing race and equity as \textit{a nervous area of government}.} Moreover, \oldaneditb{he shared that} employee uncertainty about working with a racial equity lens is exacerbated by a fundamental misalignment: City centers race in equity but \textbf{legal challenges \oldanedit{at the federal level} prevent them from pursuing tactics to directly impact racial equity} \cite{heckler2017publicly}, such as including race as a criterion to prioritize projects. This is an unfortunate limitation that perpetuates what \citet{starke2018administrative} call an "administrative racism cycle".

Despite these challenges, \oldaneditb{P36 feels} the Equity Office pursues meaningful ways to advance equity. One important component \oldaneditb{he believes has impacted} their success is the personal relationships they have developed with City employees: employees who participated in their department's equity assessment continue to reach out to them for assistance on new projects. The office also mobilizes through bi-weekly meetings with employees across departments to discuss progress and seek one another's feedback. And in the face of barriers to pursuing racial equity initiatives, the office turns to the use of "proxies" to exact change for racial equity, such as prioritizing aid to low-income communities. While not ideal, \oldaneditb{P36} explained that this is \oldaneditb{the office's} best option in the face of legal pushback \oldanedit{at the federal level}. 

\subsection{Equity in Data}\label{Findings_data}
Regardless of whether their department has formalized equity responsibilities or how long they have been practicing with an equity lens, employees expressed a high awareness about the role of data in supporting or hindering equitable decision-making. Additionally, though we noted in \ref{Findings_Context} that some participants equated equity with equality, we observed that considerations about obtaining and analyzing data for disparate impact across sub-groups still arose. 

We synthesize participants' data equity considerations into five types of data needs, detailing in Table \ref{table:dataeq} and below how each corresponds with equity impacts and additional trade-offs to keep in mind.
\begin{table}[htbp]
\footnotesize
\begin{tabular}{@{}clll@{}}

\toprule

\textbf{No.} 
& \multicolumn{1}{l}{\textbf{Types of Data Needs}} 
& \multicolumn{1}{l}{\textbf{Impact on Equity}} 
& \multicolumn{1}{l}{\textbf{Trade-offs To Keep in Mind}} \\ 

\midrule

1 & \begin{tabular}[c]{@{}l@{}}Standardized data \\ \textit{Health, Watershed,} \\ \textit{Development}\end{tabular} & \begin{tabular}[c]{@{}l@{}}(1) Discern sub-groups with different needs currently\\ generalized and overlooked when grouped in larger \\demographic groups \\(2) Identify individuals across programs or services \\(3) Improve accuracy of reporting metrics\\ around outcomes \end{tabular} & \begin{tabular}[c]{@{}l@{}}Categorization of characteristics \\such as demographics can be \\unreliable \end{tabular} \\ \midrule

2 & \begin{tabular}[c]{@{}l@{}}Granular data \\ \textit{Transportation, Watershed,} \\ \textit{Health, Development} \end{tabular} & \begin{tabular}[c]{@{}l@{}}(1) Identify patterns of inequities masked by \\aggregate data such as census tracts \\or broad classifications \end{tabular} & \begin{tabular}[c]{@{}l@{}}Privacy concerns \\Data misuses \end{tabular} \\ \midrule

3 & \begin{tabular}[c]{@{}l@{}}Centralized data \\ \textit{Housing, Small Business, } \\ \textit{Watershed, Transportation} \end{tabular} & \begin{tabular}[c]{@{}l@{}}(1) Increase accessibility of applying for \\government services \\(2) Align metrics that have interdepartmental \\dependencies and streamline inter-\\departmental service offerings \end{tabular} & \begin{tabular}[c]{@{}l@{}}Measurable equitable outcomes \\ remain difficult to define \end{tabular} \\ \midrule

4 & \begin{tabular}[c]{@{}l@{}}Representation in data \\ \textit{All} \end{tabular} & \begin{tabular}[c]{@{}l@{}}(1) Increase representation in data from minority \\groups through community engagement \\(2) Collect more qualitative data and \\lived experiences \\ \end{tabular} & \begin{tabular}[c]{@{}l@{}}Time-consuming and expensive \\to collect \\Large data overload can lead to \\analysis paralysis \\ \end{tabular} \\ \midrule

5 & \begin{tabular}[c]{@{}l@{}}Appropriate proxies to use \\ \textit{All} \end{tabular} & \begin{tabular}[c]{@{}l@{}}(1) Avoid legal issues of \\ pursuing racial equity directly \end{tabular} & \begin{tabular}[c]{@{}l@{}}Addressing proxies of racial equity \\does not equate to addressing \\racial equity \end{tabular} \\ \midrule


\end{tabular}
\vspace{2mm}
\captionsetup{width=1\textwidth}
\caption{Employees' thoughts on considerations for how to improve equity in data use and the accompanying potential trade-offs. \oldaneditb{(All impacts on equity mentioned revolved around external equity.)}}
\label{table:dataeq}
\end{table}

\subsubsection{Current Data Practice Needs for Equity}

\textbf{\oldaneditb{Departments need} standardized \& granular data \oldaneditb{to establish} reliable metrics and uncover patterns of inequities.} \oldaneditb{We heard from multiple departments that they} currently rely on insights from City data and data from official, non-city sources to determine what services to provide and to whom. However, their data sources \oldaneditb{often lack} standardization \oldaneditb{and sufficient} granularity to accurately measure whether outcomes are equitable. Health described ongoing work to standardize internal data and data provided by community partners. \oldaneditb{P8 talked about how} Health's programs currently collect the same type of information using different categories, leading to conflicting records. \oldaneditb{But once they can standardize data collection, he explained} the department will be able to track health outcomes accurately and serve intersecting program needs. Health also \oldaneditb{shared} plans to collect more granular data to support equitable outcomes: while their metrics currently reflect successful outcomes for the Asian Pacific-Islander diaspora, P9 said if this demographic group is broken down further, they may discover a sub-community that was overlooked due to being lumped in with a broader categorization.
Watershed also desires more granular data to facilitate equitable outcomes. \oldaneditb{P24 and P25} explained the Social Vulnerability Index (SVI) metric \oldaneditb{Watershed} currently uses to target aid efforts is broken down by census tracts. However, \oldaneditb{their data team} has analyzed these to find that census tracts actually mask households requiring extra support by tagging regions as uniformly low risk when they actually contain households at high risk of displacement.


\textbf{Centralized data \oldaneditb{is important for bridging} interdepartmental dependencies.} Though each department sets its own goals, measures of success, and desired outcomes, \oldaneditb{we heard employees reflect that} the data they \oldaneditb{each} collect and report on are intertwined with the decision-making and outcomes of other departments. \oldaneditb{Employees talked about how} City currently lacks centralized data for across- and within-department use, \oldaneditb{and they} viewed this as a challenge when measuring (long-term) outcomes and an unnecessary barrier for citizens trying to access government services. 

For example, P13 explained Housing's goal to provide affordable housing goes beyond house prices. Ultimately, their outlook centers around "long term affordability" which is tied to other factors, e.g., access to public transit, grocery stores, and high performing schools. Thus, \oldaneditb{they asserted} the success of Housing's equity goal is impacted by other departments, such as Transportation's decisions for where to build public transit. Similarly, Transportation explained that they focus on "placemaking", the reimagining of the "relationship between transportation, private development, and civic spaces" \cite{s_amp_me_2022}. P3 explained they consider "how someone was, what is your path getting there, and what makes you want to \textit{stay} there and feel comfortable?", \oldaneditb{a comment that suggested to our research team how} Transportation must also factor in other departments (e.g., Parks, Public Works, Housing) when pursuing equity goals.

This lack of centralized data also creates barriers for citizens accessing city services. Small Business and Housing explained that citizens must often complete time-consuming questionnaires to queue for services. \oldaneditb{Employees believe} this becomes an obstacle in serving citizens equitably as those most in need are unlikely to have the institutional trust and time for data input, abandoning forms. P13 explained the lack of centralized data also means citizens must re-enter their information for each program if they seek multiple department services.

\textbf{\oldaneditb{Data must be representative and inclusive of community voices to ensure it elevates diverse} lived experiences:} 
\oldaneditb{All departments emphasized the importance of} collecting citizen feedback to center stakeholders in department efforts. However, Development, Watershed, and Health mentioned this type of data may not be representative and is prone to self-selection bias. In the past, Watershed relied on 311 data to target efforts.\footnote{311 is a program that allows citizens to report complaints to their government.} However, P25 told us they no longer use this exclusively after realizing those most likely to use 311 speak English, have access to a phone and the time to call, and trust that City's motives. P25 acknowledged that historically, marginalized communities are least likely to call 311 due to institutional mistrust and privacy concerns \cite{white2018promises, levine2014political, brayne2014surveillance}. To improve diverse citizen representation, \oldaneditb{P22 and P24 explained} Watershed is deploying a multi-pronged community engagement effort to update their department mission. This includes launching a community ambassador program that compensates ambassadors for their time and efforts spent collecting stories of lived experiences from citizens; and creating a program to fund local non-profits to distribute surveys to marginalized communities. Prior frameworks, particularly GARE's Equity Toolkit, while not centered around data practices and equity, have similarly emphasized increasing representation by centering community voices. 

\textbf{\oldaneditb{It is often necessary to seek} appropriate proxies \oldaneditb{to help advance equity in the face of restrictions} such as legal limitations:} Just as the Equity Office shared about using proxies to circumvent federal legal challenges for pursuing racial equity, other departments mentioned \oldaneditb{finding available, appropriate} proxies as a data need for equity. \oldaneditb{P6 from} Development \oldaneditb{pointed out their use of} gaps in City's tree canopy, which align with historical patterns of racial disparities, to direct their efforts. Watershed \oldaneditb{employees described} using a socioeconomic indicator called Social Vulnerability Index (SVI) \oldaneditb{from the United States Center for Disease Control (CDC)} \cite{agency2021glance}, which includes race amongst other factors to identify flood-risk regions for targeting assistance efforts. \oldaneditb{P11 and P12 from} Housing \oldaneditb{relayed department attempts} to incorporate race as a criterion for allocating displacement prevention funds to minorities, \oldaneditb{but} ultimately replaced it with low-income metrics at the advice of City's legal counsel.

\subsubsection{Trade-Offs Considerations for Data and Equity}
\oldaneditb{Participants were also conscientious about the dilemma that in some cases, addressing the data needs they discussed could lead to other challenges to overcome before being able to use the data for equitable outcomes.}


\textbf{Mitigating privacy, data misuse, and unreliable standardization.} 
Even as they shared a desire for more granular, standardized data, employees raised concerns about privacy and misappropriation of this data. For example, Transportation explained that in an attempt to increase accountability and civic engagement, they shared anonymized micromobility movement data (scooters, ebikes) on City's open data portal. However, they removed this after a researcher approached the department to show he could deanonymize individuals. Development employees emphatically disapproved of ever collecting citizen demographics even if it could support equity. \oldaneditb{P5} worried about a future where those in power could use this data to discriminate. 
P5 and P6 \oldaneditb{also} shared concerns about the reliability of collecting demographic data and asking people to self-identify. \oldaneditb{We reflect that} this raises an important point around race dimensions which has been explored by researchers such as \citet{hanna2020towards}: though increased granularity can reveal unique experiences of different groups, self-identification is very personal and subjective, and may not be appropriate for every outcome City intends to use it for. 

\textbf{Mitigating analysis paralysis.} 
Collecting representative data is only one component for advancing equity. Departments still face challenges in making sense of the overwhelming quantity of information collected. Health recently completed a comprehensive health inequities survey with members of their entire department separate from the Equity Office's equity assessment. However, they are currently in a holding pattern as they wait to hire an external consultant to help them analyze the results. 

\textbf{Recognizing the limitations of proxies.} While they have brainstormed creative alternatives to use for racial equity, employees were quick to point out how proxies can overgeneralize and fail to capture the nuances of race. A few were particularly frustrated about the insufficiencies of proxy metrics: P12 shared that excluding race ultimately fails to address racial equity, saying "[if] that's the case, and we're using geography and all of these other proxies, there are some people who experienced harm who are never repaired."

\subsection{Equity in Technology Applications}\label{Findings_tech}
%
When discussing technology use and equity, employees often described concrete use cases. \oldaneditb{These were important in clarifying for our team the equity goals employees have in mind when they are deciding to adopt a tool and the corresponding} boundaries and functionalities \oldaneditb{they believe} government technology use \oldaneditb{should respect}. In Table \ref{table:techeq}, we summarize these ideas about the boundaries of acceptable technology along with equity goals---expressed as "Impact on Equity"---and 1-2 corresponding use cases. The table examples are intended to be illustrative and not exhaustive of every instance we heard from employees. 

\oldaneditb{At a glance, some boundaries may seem deceptively obvious ("Automate Objective Tasks"). But viewed alongside use cases employees surfaced and how these advance equity efforts, we believe this can help researchers recognize real-world needs and impacts, ensuring they do not overlook such boundary suggestions.}
Below, we expand on a use case for each boundary to help \oldaneditb{demonstrate timely pursuits for government technologies situated in employees' experiences.}
\begin{table}[htbp]
\footnotesize
\begin{tabular}{@{}clll@{}}

\toprule

\textbf{No.} 
& \multicolumn{1}{l}{\textbf{Boundaries on Tech}} 
& \multicolumn{1}{l}{\textbf{Impact on Equity*}}
& \multicolumn{1}{l}{\textbf{Use Case Example}}  \\ 

\midrule

1 & \begin{tabular}[c]{@{}l@{}}Inform Decision-Making\\ \textit{CIT, Innovation, Watershed} \end{tabular} & \begin{tabular}[c]{@{}l@{}}\textbf{Internal:} Assist with resource allocation \\ \textbf{External:} Identify groups of people who \\may have been overlooked in terms of \\achieving equitable outcomes.\end{tabular} &\begin{tabular}[c]{@{}l@{}}(1) Analyze data community survey\\ feedback or other city data to inform \\human decision-making \end{tabular} \\ \midrule


2 & \begin{tabular}[c]{@{}l@{}}Automate Objective Tasks\\ \textit{Development, Health, Housing,} \\ \textit{Small Business} \end{tabular} & \begin{tabular}[c]{@{}l@{}}\textbf{Internal:} Allocate employee time from \\manual, objective tasks to responsi-\\bilities such as engaging intentionally \\with equity trainings \\\textbf{External:} Lower cost and barriers to \\accessibility to serve citizens more \\inclusively \end{tabular} & \begin{tabular}[c]{@{}l@{}}(1) Automate manual reports or data \\collection\\ (2) Allow citizens to complete portions \\ of inspections such as measuring their \\space with their phone\end{tabular} \\ \midrule

3 & \begin{tabular}[c]{@{}l@{}}Improve City Presence\\in the Community\\ \textit{Health, Watershed} \end{tabular} & \begin{tabular}[c]{@{}l@{}}\textbf{External:} Cultivate more inclusive \\and accessible spaces for historically \\underrepresented groups\end{tabular} & \begin{tabular}[c]{@{}l@{}}(1) Hybrid meeting technologies to \\ invite/include more people to the table \\(2) Help departments be more present \\and hear from more of the community \end{tabular} \\ \midrule

4 & \begin{tabular}[c]{@{}l@{}}Remove Subjectivity\\of Municipal Processes\\ \textit{Housing, CIT} \end{tabular} & \begin{tabular}[c]{@{}l@{}}\textbf{Internal:} Improve equitable outcomes \\ for performance reviews of employees\\\textbf{External:} Reduce human biases and \\ open up spaces that were previously \\ dependent on generational knowledge \\ and political influence\end{tabular} & \begin{tabular}[c]{@{}l@{}}(1) Detect human bias in performance\\ reviews\\(2) Streamline processes that exhibit \\human bias such as event permitting\end{tabular} \\ \midrule

5 & \begin{tabular}[c]{@{}l@{}}Improve Public Safety\\ \textit{Fire, Transportation}\end{tabular} & \begin{tabular}[c]{@{}l@{}}\textbf{External:} Provide services to typically \\unsafe and inaccessible hazard-\\ous situations\end{tabular} & \begin{tabular}[c]{@{}l@{}}(1) Expansion of air-, ground-, and \\water-based solutions \\ (drones, robots, boats)\end{tabular} \\\midrule


6 & \begin{tabular}[c]{@{}l@{}}Clarify Government Processes\\ \& Improve Information Retrieval\\ \textit{Development, Health, Housing,} \\ \textit{Watershed} \end{tabular} & \begin{tabular}[c]{@{}l@{}}\textbf{External:} Lower barriers that currently \\gate-keep individuals without insti-\\tutional knowledge about how to \\ influence city government\end{tabular} & \begin{tabular}[c]{@{}l@{}}(1) Intuitive tools to help citizens \\retrieve city's services with more ease\\(2) Improve website tools to support \\ understanding of council processes \\ and facilitate civic engagement\end{tabular} \\ \midrule


\end{tabular}
\vspace{2mm}
\captionsetup{width=1\textwidth}
\caption{Boundaries identified by employees for acceptable uses of technology to ensure equitable outcomes. \\ \oldaneditb{*"Internal" refers to internal equity impacts on City's employees. "External" refers to external equity impacts on City's communities.}}
\label{table:techeq}
\end{table}

\subsubsection{Considerations for Technology Use to Advance Equity}
\hfill \break

\textbf{Inform Decision-Making.} 
\oldaneditb{Past research has focused on public sector decision-making tools as it relates to high-stakes contexts and making predictions \cite{saxena2021framework} in part motivated by the desire to mitigate human biases. Conversely, we observed desires from departments for informing decision-making by making datasets more digestible to expose (and counter) their own biases or as P25 put it, "blinders that I have on that I don't know I have on".} \oldaneditb{This falls in line with \cite{zhang2023deliberating}'s findings that participants wanted AI tools that act as "bias checkers" to help individuals challenge personal biases during decision-making.}

\oldaneditb{To overcome these "blinders", they imagined tools to help analyze and identify which communities they still lack input from when designing equitable initiatives.} Referencing Watershed's ongoing department-wide community engagement effort, P24 wanted a way to filter through survey responses and identify gaps on what community groups still need to be heard from, suggesting a dashboard to display these insights. \oldaneditb{He felt this would} support external equity by ensuring the voices of marginalized groups are heard. 

\begin{quote}
    "There's probably some role that AI or technology could play in helping us be more present and hear from the community, also, \textit{to make sense of all the different things that they're telling us}...you automate pulling [the data] all together, but then like, the kind of the decisions and the work and thinking about it still is done by humans." ---P24
\end{quote}


\textbf{Automate Objective Tasks.} \oldaneditb{Though automation of objective tasks may seem a rudimentary boundary, the use cases from employees highlight on-going issues that have not been addressed well yet. Some use cases can also help explicate practitioner and department-level workflows that may not be known to researchers. \oldaneditb{For example, Development described a scenario about inspectors' responsibilities to explain why automating some can have an equity impact on community members. This revealed details about the permitting process and inspectors' tasks that we previously did not know---expanded below---that the use case helped elucidate.}}

P5 from Development explained that one of the biggest expenses for citizens requiring the department's services is the inspector conducting on-site visits. He \oldaneditb{gave the example about how permits and on-site visits are currently required when citizens are installing water heaters. He} suggested \oldaneditb{instead} the use of software to automate inspectors' objective tasks, such as room measurements, so that citizens can use their own devices to complete them. This would support external equity in two ways: lower the cost for citizens needing inspection services and provide inspectors with more time to dedicate to subjective tasks that require interpretation of complicated code---\oldaneditb{such as interpreting City code to resolve a citizen's request to waive an unreasonable \$4k fee}.

\begin{quote}
    "That's the biggest cost of the permit---like the inspection piece, like somebody having to physically get out of their desk. [Inspectors] can only do so much in an hour...Maybe it's a customer logs in, they hold their cell phone camera up, they walk around...and it kind of creates like that blueprint of that space [with] LIDAR data?" ---P5
\end{quote}

\textbf{Improve City Presence in the Community.} \oldaneditb{Participants in Health and Watershed showed a particular awareness around upstream factors that impact equity and addressing these to effectuate meaningful change, such as the need for thoughtful, inclusive modes of participation in civic engagement. We acknowledge a caveat to the extent on equity is how well the feedback is gathered, synthesized, and ultimately incorporated into municipal processes and actions. However, participants' ideas are helpful insights rooted in their experiences working with communities.}

P9 from Health expressed approval for technology that further supports hybrid meetings. She explained how since the pandemic, she now believes "if we are inviting someone to the table, and it's not a hybrid option, I don't think that invitation is genuine". She explained how they are currently using this technology to support internal equity and inclusion of employees and have an interest in using it for external equity to improve outreach efforts with the community. This echoes what we heard in our initial interview with Housing about their use of software to run multi-lingual and multi-modal virtual townhalls. Relatedly, P10 from Health also discussed a desire to set up community engagement centers situated in neighborhoods of under-served populations to supporting hearing from these groups they historically lack input from. She envisioned this as computer labs for citizens to use to give their feedback and potentially even obtain (health-related) training from.

\textbf{Mitigate Subjectivity of Municipal Processes.} 
\oldaneditb{}
While employees were hesitant about AI biases, some expressed interest in technology to mitigate \textit{human} biases. \oldaneditb{Uniquely, employees highlighted (1) a key difference is these tools they desire are not data-driven, whereas many public sector technologies are data-driven; and (2) intertwined with mitigating human biases is reducing equity barriers stemming from generational know-how or political influence.

For instance,} P12 explained why it is important for event permitting \oldaneditb{(e.g., to hold cultural festivals)} to be streamlined: confusing municipal processes like this can exacerbate inequities, especially for people of color, because they can require proposal presentations in front of intimidating committees or knowing the right people and right policies to get approvals. 
\begin{quote}
    "This is how, to me, BIPOC communities lose. Because they don't have \textit{relationships}, they don't have \textit{phone numbers}, they don't have access to the people who are \textit{in control}...if you don't have a clear policy, policy and procedures, then you can't follow it. And as a result, those people without connections will lose every single time." ---P12
\end{quote}




\textbf{Improve Public Safety.} \oldaneditb{Two departments expressed interest in technologies for advancing public safety, such as} P16 from Fire who wanted increased drone, robot, and boat use. In his experience, these tools have helped Fire gain situational awareness of hazardous situations not previously accessible. He gave an instance where Fire used a drone for a suicide \anedit{rescue} mission to assess if the individual had weapons and decide how to safely deploy rescuers. We note that during this interview, both benefits of drones and concerns---e.g., the misuse of drones for protest surveillance---emerged. 

\begin{quote}
    "People end up climbing these tower cranes that we have in our city. And before we send up our rescuers, we can get up there pretty quickly, make sure they don't have any weapons on them...Learn if they're going to be a danger to our people. And then we can be able to, to gauge what our next steps are for for our personnel, how we can better help, you know, the people who are in distress. But yeah, that's happening - that actually happens quite a bit." ---P16
\end{quote}



\textbf{Clarify Government Processes and Improve Information Retrieval.} \oldaneditb{Multiple employees expressed frustration over City's website being unnecessarily difficult for citizens to search for services.} Even if they find relevant information, P6, P11, and P24 noted citizens have to navigate "legal-ese" to make sense of things such as city council processes or land development code requirements. Technology to improve information retrieval and make government processes more intelligible to citizens can support external equity by lowering the barriers for citizens to navigate municipal processes.

\begin{quote}
    "There's a lot of information that we have that people want to know. But they have a hard time finding it on our website, which is terrible. And they have a hard time navigating the council process...I'm a professional planner, and it's really hard...there might be some opportunity to...take this really complicated information that exists in code and criteria and legal-ese and translate it into something people can understand." 
    ---P24
\end{quote}

This challenge of navigating niche municipal processes echoes \citet{kim2024public}'s \anedit{description of} small business owners' experiencing \textit{knowledge inequities} while trying to obtain necessary building permits to set up their storefronts.

\subsubsection{AI-Specific Concerns}
Although our focus on government technology and equity considerations was around general technology use, we were curious about whether employees had specific feelings around departmental adoption of AI for equity goals. When asked, some employees used the opportunity to imagine their ideal AI tools while others strongly opposed AI.

Those concerned linked their hesitations to City's lack of infrastructure to evaluate AI, disapproval of replacing human decision-making, and unease around the ethics of AI. (All employees were adamant against any replacement of human decision-making and customer service roles.) The most vehement responses against AI came from employees of Equity, Housing, and Innovation. P36 worried City did not have the capacity yet for AI tools, risking more harm than good. P12 expressed wariness and privacy concerns around AI \oldaneditb{and} advancements in technology in general, nervous that people in power may use it unethically. P32 explained that as a local government, technology must be unequivocally beneficial so that citizens are not treated like "guinea pigs" which they do not believe AI currently is.

Employees also shared problematic practices they have observed about vendors selling AI to City. P16, who operated AI for Fire, was frustrated that retailers only care about pushing products without asking customers for feedback on how to improve them. Innovation was skeptical over the intentions of vendors: "municipal spaces [are] being seen as a space to, to develop and to sell these technologies, whether or not we need them, or whether or not that they're an appropriate match", such as one seller pitching them AI to use with people experiencing homelessness. P32 explained vendors set unrealistic expectations and "capture the imagination" of elected officials or staff. P33 added that "most cities can’t...assess the structure of a technology that’s being pitched to us" for equity implications. 

\section{Discussion} 
Our findings expand CSCW \oldaneditb{research} on designing equitable government technology and data use: \oldaneditb{through interviews with a city government's employees, we focused on understanding their equity-specific practices and contexts, which we then situated to surface corresponding data and technology perspectives}. We discuss two areas for future work based on our results. First, we examine what it means to foreground equity in data and technology and the implications for design (Section \ref{discussion1}). Second, we reflect on how to support public sector employees towards meaningful use of tech and data in the context of equity (Section \ref{discussion2}).


\subsection{Foregrounding Equity in Government Data Use and Technology Design} \label{discussion1}
\oldaneditb{We advocate for future research to embrace} creating space for participants to delve specifically into their practices and perspectives around equity. \oldaneditb{This is important to advance data and technology for equitable public administration for two reasons. First, without this focus, participants may hesitate to share ideas} they perceive being incompatible with competing public administration principles. Recall P10's idea for Health to set up neighborhood centers in under-served areas as accessible sites for community training and input gathering. Prior to our session, she had never shared this idea with her colleagues for how it may be too expensive and time-consuming (i.e., infringing the public administration pillars of \textit{economy} and \textit{efficiency}). Second, we realized that beginning and re-centering discussions around equity challenges and desires helped \textit{us as researchers} recognize participants’ ideas that we may have overlooked as inconsequential or not considered due to disciplinary differences. To illustrate the value of foregrounding equity practices, we expand on two ideas from our findings for CSCW scholars to pursue for equitable public sector data and tech use.
\subsubsection{Creating tools \& data processes that support long-term equity}


Typically, researchers have been concerned with  fairness metrics to measure and mitigate biases of tech systems and AI. These usually center around assessing whether such outcomes or decisions are fair and unbiased \cite{mehrabi2021survey, dwork2012fairness} \oldaneditb{with popular fairness metrics closely aligned to equality---equal outcomes---as opposed to equity}. \oldaneditb{In our study though,} several employees raised concerns about how to collect data on and measure for \oldaneditb{equity, particularly} \textbf{long-term equity}. \anedit{For example,} P13 from Housing viewed affordable housing as not just about house prices but whether recipients can access \oldaneditb{related essential} services, \anedit{including} those provided by other departments like Transportation or Health, to improve \oldaneditb{and} maintain their quality of life over time. Existing metrics will be insufficient, and assessing true long-term equity outcomes will require creating new measures that also consider inter-departmental outcomes and equity goals. This is a valuable line of inquiry for employees and designers to collectively deliberate over to advance sustainable, equitable systems and tools. 

\oldaneditb{One potential path for creating short-term and long-term equity measures is the framework outlined by \citet{jacobs2021measurement}. The authors explain how unobservable constructs like fairness must be inferred using observable properties. This is akin to our participants' use of targeting \textit{proxies} to advance equity which is also an unobservable construct. The authors draw on quantitative social science metrics of \textit{validity} and \textit{reliability} as a foundation for proactive auditing that surfaces mismatches in how unobservable constructs are being operationalized versus measured. Building on \citet{jacobs2021measurement}'s framework for fairness metrics, future research for short-term and long-term equity metrics could make a meaningful attempt to quantitatively assess our participants' early ideas about long-term equity for opportunities to improve these concretely in a model. \anedit{For example, a model to determine locations for} affordable housing that effectuates long-term equity outcomes \anedit{could include these} initial variables \anedit{participants raised}: cost (of housing), conditions (quality of housing and responsiveness to tenant concerns), obtaining and accessing a job (proximity to public transportation), and food needs (proximity to grocery stores). Alternatively, researchers might look to \anedit{create} tools that help government employees critically and quantitatively evaluate the assumptions of the constructs they design \anedit{using} reliability and validity metrics.} 

\subsubsection{Bridging siloed data across departments to support service delivery and discovery for constituents}

Some researchers have suggested that digital systems sufficiently address routine tasks (e.g., cross-departmental data sharing) \cite{saxena2024algorithmic, veale2019administration}, which has led to decision-makers focusing on automating "high-stakes decision-making". \oldaneditb{However, efforts around this have surfaced concerns that datasets being used actually amplify historical inequities \cite{gillingham2019decision, eubanks2018automating, tolan2019machine}. To address that, practitioners have pointed out their need for granular data to assess system fairness \cite{madaio2022assessing} and "fairness-aware datasets"---datasets that have sufficient representation for bias detection \cite{holstein2019improving}.} 
\oldaneditb{In response, researchers often push for disaggregated demographic data to improve datasets for fairness and equity \cite{kauh2021critical, kim2024public}. 
Yet \citet{kim2024integrating}'s recent work and our findings surfaced concerns that collecting constituent demographics can actually backfire (e.g., P5 was worried citizen demographic data could be used by an authority figure to actively prevent specific racial or ethnic groups from obtaining development permits).}

\oldaneditb{On the other hand}, our participants made clear \oldaneditb{they face more urgent equity-related data challenges} in City: they still lack and desire technical infrastructures to accomplish \oldaneditb{the aforementioned routine tasks}. For example, \oldaneditb{they still need} tools to centralize interdepartmental data. \oldaneditb{This could assist with upstream equity efforts by lowering duplicative efforts of citizens applying for services across city departments, and even enable personalized recommendations to citizens about new services they may qualify for. This can also assist with downstream equity efforts: \anedit{streamlined} interdepartmental data collection and sharing would help employees around long-term equity which \anedit{encompasses} outcomes related to different departments (e.g., whether recipients of affordable housing are also able to access transportation to reach workplaces)}. 

We \oldaneditb{centered equity contexts in our study by probing employee equity goals and practices} separate from technology and data considerations before integrating those two topics back in. \oldaneditb{While seemingly simple, we} imagine that researchers creating toolkits or resources for public sectors can foreground employee equity contexts in similar ways \oldaneditb{to surface meaningful discussions}. For example, a group using \citet{kawakami2024situate}'s Guidebook to deliberate over AI adoption might precede it with a frank discussion amongst employees about how they interpret and practice equity in their work.
\oldaneditb{To clarify, we do not suggest these discussions should only occur in conjunction with toolkits or guidebooks}. \oldaneditb{We simply advocate that} starting with a discussion about their equity practices and goals separate from a proposed technology can be a more accessible starting point for employees before they deliberate over more daunting, high-stakes questions such as, "Can we agree on a definition of fairness and equity in this context?" \cite{kawakami2024situate}. 

\subsection{Considerations On How to Support Employees in Operationalizing Equity}\label{discussion2}




A number of city governments have initiated efforts to create equity offices, special task forces, and initiatives in the hopes these can address concerns around technology inequality \cite{kimmerling_kim_2022, tomer_fishbane_2022, nyctaskforce_2019, nycCTO_2021, sfo2019}.
%
%
\oldaneditb{To learn best practices and establish their own office, the City and County of San Francisco commissioned a report in 2019 to evaluate the characteristics of 33 cities' Equity Offices. This found that equity mission statements differ widely, 
support from high-ranking authorities for visibility and accountability was an important factor for success, and no "published criteria for evaluating results or effectiveness in creating a new office of equity" was discovered \cite{sfo2019}.} 
\oldaneditb{In addition to limited criteria existing for efficacy measurements, there is also limited work that comprehensively documents how equity offices have impacted equity practices since their conception.}
\oldaneditb{Our findings situated in a city with an Equity Office offer a brief glimpse into how this office has tried to impact equity practices across different departments. Based on this, we share initial thoughts about how to support employees in operationalizing equity.} 

\subsubsection{Establishing a centralized Equity Office or Department.} One finding that surprised us was the limited formal authority that the Equity Office held for enacting change. This central body was intended to efficiently spearhead and oversee equitable outcomes throughout City's departments. Yet, both its existence and its weakened authority actually appeared to be contributing factors to participants' sometimes unclear awareness around their accountability towards practicing equity. 

\oldaneditb{This leads us to reflect that while} a centralized department to formalize and oversee best practices and standards makes sense in theory,
in practice, \oldaneditb{it} could still face challenges if lacking official authority or if operating with a small staff. 
Additionally, employees we spoke to described some colleagues who were nervous about messing up equity themselves, preferring to be told what to do. A department specifically to oversee equity or technology equity could contribute to further uncertainty amongst employees regarding whether they must personally ensure equity in their work or if the centralized department will take care of it for them. \citet{gooden2011advancing} warn that treating social equity as a stand-alone topic instead of weaving it throughout public administration curriculum like its counterparts---economy, efficiency, effectiveness---dilutes its comparative importance. \anedit{In accordance,} governments must take care to ensure that having an equity department \oldaneditb{or} office does not create the misconception that equity is not a responsibility for the select few, but \anedit{instead makes clear that equity} must be practiced by \textit{all} employees.

This does not mean a centralized department cannot be beneficial. Based on employee experiences, there are strategies to mitigate some of the drawbacks mentioned. For example, P36 from Equity described how in lieu of formal authority, the Equity Office depends heavily on relationships they have with employees across departments to organize efforts for equity in everyday responsibilities. Formalizing these cross-departmental roles and relationships can be one method for strengthening the influence of a nascent equity oversight body and countering the misconception of equity as an add-on. 

\subsubsection{Practicing equity without an official equity body.}
\oldaneditb{In some instances, cities may wish to practice equity but lack the resources to stand up an official office. We give two ideas to consider in the meantime. First, as pointed out above \textbf{the relational aspect of practicing equity can be harnessed by municipal governments that do not have capacity for a central equity body}.}
\oldaneditb{One path forward can be establishing a working group comprised of representatives from each department. Through regular meetings, members can collectively tackle on-going projects and establish a code of best practices. We suggest a first step for these groups is establishing alignment on how employees define and practice equity. As found by \cite{sfo2019}, the scope emphasized within equity missions across cities varies (e.g., race only, race and social, race and disability). Our participants similarly emphasized a range of characteristics. Relatedly, our findings and those of \citet{kim2024public} surfaced persisting misalignment in the concepts of "equity" and "equality". Aligning on these would be necessary to meaningfully practice equity.
}

\oldaneditb{Second, we echo the importance stated by \citet{kim2024public} for government employees to engage with constituents early on in the technology design process to avoid a misalignment between citizen needs and technology deployed. Even without an equity office, employees can pursue meaningful community engagement within their departments.  This also aligns with our participants’ desires for technology that improves their community engagement efforts for advancing department equity goals. In sum, to guide their efforts for equitable outcomes, employees should seek ways to center early and sustained community engagement throughout their projects and initiatives.}

\oldaneditb{
In contemplating how to support employees without an equity office, we are reminded by \citet{newman2023experience} that many official efforts for DEI, including formalized equity offices, have been scaled back in recent years due to turnover of administrations and each one's motivations. Finding unofficial ways---such as this working group---to push forward equity separate from formal equity initiatives or positions while still incentivizing personal employee accountability will be crucial to ensure past efforts are not for naught. 
}

\subsubsection{Who shoulders the burden of upholding equity work.} Our final consideration \oldaneditb{for supporting employees in operationalizing equity is examining} who the burden of upholding ethics or equity falls on. \citet{rakova2021responsible} and \citet{wong2021tactics} discuss how the work of fair AI is often shouldered by individuals who personally recognize the importance of the work. We noticed in City that equity is similarly carried by specific individuals or departments. Innovation commented how the Equity Office sees them as partners in helping advance equity with other departments, and Small Business talked about the pressures of having to nudge other departments to think more about diversity, inclusion, and bias. Echoing \citet{madaio2022assessing}'s caution about how dependence on diverse practitioners for fair AI can exacerbate epistemic burden \cite{pierre2021getting}, how can equity interventions ensure robust oversight into technology creation, procurement, and evaluation processes, while not forcing specific groups to take up the mantle on their own? One way could be through providing time off or linking equity with employee responsibilities formally as Health was doing, similar to a suggestion raised by \citet{rakova2021responsible} and \citet{wong2021tactics} to tie equity and fairness metrics to worker evaluations. 

\section{Limitations} 
Our interviews were with U.S. city government employees. We recognize that there are differences in how governments operate across cities as well as at the city, state, federal, and international levels. The city that we studied was in an early stage of AI \oldaneditb{and} ML adoption. \oldaneditb{Additionally, we cannot be sure whether the presence of an equity office had a broad influence on participants' responses. For instance, if City did not have an equity office, would the responses to our questions have resulted in a lower awareness about department-level equity goals or even a lack of equity goals? While we cannot be sure, we believe even in the absence of an equity office or department, there might be less agreement over the concept of equity, but given equity is a pillar of public administration, employees would likely still be grappling over it. In fact, there may be higher instances of epistemic burden placed on individuals with naturally higher awareness of practicing equity.} We acknowledge that \oldaneditb{including City Equity Office employees} in the first round of interviews, and each interviewee's comfort in discussing equity or technology \cite{gooden2015race} may have impacted the insights employees felt comfortable sharing. We also acknowledge potential bias in participant selection due to factors such as availability, topical interest, and degree of familiarity with the Equity Office who circulated the call for participation in first round interviews. Finally, future research should be done with other city practitioners in different contexts using diverse methods such as interviews, observations, or surveys.

\section{Conclusion}
In our paper, we investigate\oldaneditb{d} how \oldaneditb{employees of a} local government pursue equity goals to situate equity considerations within their data and technology practices. %
We interviewed thirty-six employees from ten departments in a U.S. city government. We describe the equity contexts that shape their work and challenges they face, followed by the equity considerations employees raised in their data practices and the design space they identified for acceptable technology to advance equity.  %
From our findings, we discuss what it looks like to foreground equity in public sector technology and data use, \oldaneditb{suggesting future research directions to support government employees in tools that support long-term equity goals and metrics as well as bridging siloed data across departments to support equitable service delivery to constituents}. We also discuss practical considerations for supporting government employees in equity awareness and accountability so they can pursue equity in tech and data practices meaningfully.
\begin{acks}
\anedit{Thank you to our participants for sharing their experiences and reflections, and for their continued care and dedication as public servants. We are also grateful to our anonymous reviewers whose considerate feedback helped us improve our manuscript. This research was partially supported by the following: the National Science Foundation IIS-1939606, CCF-2217721, CNS-2313104, and DGE-2125858 grants; Good Systems, a UT Austin Grand Challenge for developing responsible AI technologies\footnote{https://goodsystems.utexas.edu}; and UT Austin’s School of Information.} 
\end{acks}


\bibliographystyle{ACM-Reference-Format}
\bibliography{references.bib}
\end{document}